\documentstyle{mn}

\title{Evidence for a warm ISM in the Fornax dEs FCC046 and FCC207}

\author[S. De Rijcke, W.~W. Zeilinger, H. Dejonghe \&
  G.~K.~T. Hau]{S. De Rijcke$^{1,*}$, W.~W. Zeilinger$^{2}$,
  H. Dejonghe$^1$ and G.~K.~T. Hau$^{3}$ \\ $^1$ Sterrenkundig
  Observatorium, Ghent University, Krijgslaan 281, S9, 9000 Gent,
  Belgium \\ $^2$ Astronomisches Institut, Universit\"at Wien,
  T\"urkenschanzstrasse 17, A-1180 Wien, Austria \\ $^3$ ESO, Alonso
  de Cordova 3107, Santiago, Chile \\ $^*$ Postdoctoral Fellow of the
  Fund for Scientific Research - Flanders (Belgium)(F.W.O)\\ Based on
  observations collected at the European Southern Observatory, Chile
  (ESO Large Programme Nr.~165.N-0115)}
\begin{document}

\maketitle

\begin{abstract} 
We present H$\alpha+$[N{\sc ii}] narrow-band imaging of FCC046 and
FCC207, two dwarf ellipticals (dES) in the Fornax Cluster. B-R and B-I
color-maps clearly show signs of dust-absorption in FCC207. FCC046 has
a very bright blue nucleus, offset by about $1.1''$ with respect to
the outer isophotes. FCC046 was classified as a non-nucleated dE4 so
the presence of its nucleus came as a surprise. Moreover, FCC046 shows
a pronounced lopsided shape. Given that FCC046 is an isolated galaxy,
it is unlikely that an encounter caused this asymmetry. The emitting
regions differ considerably between the two galaxies. Whereas FCC207
has only one central emission region, FCC046 also contains fainter
emission regions. Based on broad-band colours, its disturbed shape and
its very bright nucleus, FCC046 is akin to the class of amorphous
dwarfs. The central emission regions of both galaxies are barely
resolved under seeing conditions of FWHM$\approx 0.8''$ and we
estimate their diameters at about 60~pc. Their H$\alpha$ luminosities
can be explained as photo-ionisation by post-AGB stars in an old
population. Some of the fainter extended emission regions in FCC046
are resolved and have diameters of the order of 50$-$150~pc and
H$\alpha$ luminosities of the order of 10$^{30}$~W, comparable to
supernova remnants or nebulae around Wolf-Rayet stars. Hence, FCC046
is clearly undergoing star-formation while for FCC207 the case is not
as clearcut. We estimate the mass of the H{\sc ii} gas in FCC046 at
$M_{\rm H{\sc ii}} = 40-150 M_\odot$ (for $T=10^4$~K,
$N_e=1000$~cm$^{-3}$). The ionised-gas content of FCC207 is somewhat
higher~:~$M_{\rm H{\sc ii}} = 60-190 M_\odot$.
\end{abstract}


\section{Introduction}

Dwarf ellipticals as a rule are pressure-supported objects,
characterized by very low rotation velocities compared to their
velocity dispersions (fast-rotating dEs do exist but they are rare,
see De Rijcke {\em et al.} \cite{der}). There is currently a number of
models in vogue that attempt to explain this apparent lack of rotation
as a result of significant mass-loss. According to the ``wind-model'',
proposed by Dekel~\&~Silk \cite{ds}, dEs form from average-amplitude
density fluctuations. Most, if not all, of the ISM is subsequently
blown away after it has been heated to velocities that exceed the
galaxy's escape velocity by the first burst of supernova
explosions. This dramatic mass-loss causes a more anisotropic orbital
structure and makes the galaxy puff up. A more sophisticated version
of this scenario can be found in Mori {\em et al.} \cite{more} who
discuss the chemodynamical evolution of a $10^{10}M_\odot$ dwarf
galaxy. The first supernovae expell a supersonic outflow of gas from
the center of the galaxy. Stars form in this expanding shell and
subsequent supernova explosions further accelerate the expansion of
the shell and enrich it with metals. This model explains the outward
reddening of dEs as a metallicity effect and reflects in the
characteristic exponential surface-brightness profile.

Other scenarios take into account the fact that dEs are found
predominantly in high-density environments such as groups and
clusters. Mori~\&~Burkert \cite{mor} argue that ram-pressure stripping
is able to completely remove the gas from a dE less massive than $10^9
M_\odot$ within a few $10^8$ years. More massive dEs might be able to
retain some gas in the central region. Moore {\em et al.}  \cite{moo}
examine the role of tidal interactions between small spirals and giant
cluster-members to produce dE-like objects. On its orbit through the
cluster, a small galaxy is subjected to collisional shocks induced by
other galaxies that tear large tidal tails off it, draining angular
momentum from the remaining gas and stars. The perturbations heat the
dwarf galaxy, raising the velocity dispersion, while the mass loss
makes it inflate. Torques exerted by the collisions drive gas to the
center were it is consumed in a starburst. This process could explain
the density spike in so-called nucleated dEs.

Independent of which scenario is correct, one would expect that not
much of an ISM, if any, is present in dEs and hence that they are not
actively forming new stars. None the less, there is a growing amount
of data that demonstrates that at least some dEs have been able to
retain a sizable amount of dust and both cold and warm gas, and that
some are forming stars. We give a few examples of recent detections of
star-formation or of an ISM in dEs without attempting
completeness. Young~\&~Lo \cite{yo1,yo2,yo3} detect H{\sc i} 21~cm
emission and CO emission in virtually all Local Group dwarf
spheroidals they examined (Leo~A, NGC147, NGC185, NGC205, Sag DIG,
LGS~3 and Phoenix). NGC185 shows H$\alpha$ emission in the form of a
central extended emission region of 50~pc in diameter, probably a SNR.
On the other hand, NGC205 is devoid of emitting regions. Hence, the
presence of an ionised ISM in dEs should not be taken for
granted. NGC205 and NGC185 also show a few dust patches. The dE
A~0951$+$68, in the M81 group, posesses a high-excitation H{\sc ii}
region (Johnson {\em et al.}  \cite{joh}). The observed extended blue
light is regarded as evidence of a recent star formation
event. NGC4486A, a relatively bright dE seen almost edge-on, contains
a stellar and dust disk (Kormendy {\em et al.} \cite{kor}),
reminiscent of the nuclear disks of spiral galaxies.

Sandage~\&~Brucato (Sandage~\&~Brucato \cite{san}, see also e.g. Quill
{\em et al.}  \cite{qui}, Noreau~\&~Kronberg \cite{nor}, Marlowe {\em
et al.}  \cite{mar1,mar2}) coined the name ``amorphous dwarfs'' for
dwarf galaxies that have a disturbed appearance due to recent star
formation and the presence of dust but are not irregular enough to be
classified as Im. Marlowe {\em et al.} \cite{mar2} argue that BCDs,
H{\sc ii} galaxies and amorphous galaxies are actually all members of
the same class of star-forming dwarfs and owe their names mostly to
the selection criteria involved. Most amorphous dwarfs in the sample
of Marlowe {\em et al.} \cite{mar1} have a two-component surface
brightness profile~:~an exponential envelope and a bluer core
component. Their amorphous dwarfs show strong H$\alpha$ emission
($L_{{\rm H}\alpha} \approx 10^{33}~$W). These authors argue that it
is possible -- at least in principle -- that the cores and envelopes
of BCDs and amorphous dwarfs will fade and reach an end-state similar
to present-day nucleated dEs after they have used up their gas supply
and star-formation has ended. However, star-formation in dwarf
galaxies probably takes place in a series of mild star-bursts that
deplete the gas rather slowly. Hence, dEs must have had ancestors that
evolved more rapidly. The fact that dEs are found predominantly in
clusters while BCDs are remarkably scarce in these high-density
environments (Salzer \cite{sal}) might hold a clue~:~repeated
encounters with giant galaxies and ram-pressure stripping may have
sped up the gas-depletion process. It is clear that the present-day
star-formation rate (SFR) and ISM-content hold important clues to
understand the origin of dEs.

\begin{figure}
\vspace{8cm} \special{hscale=70 vscale=70 hsize=700 vsize=230
hoffset=-71 voffset=-240 angle=0 psfile="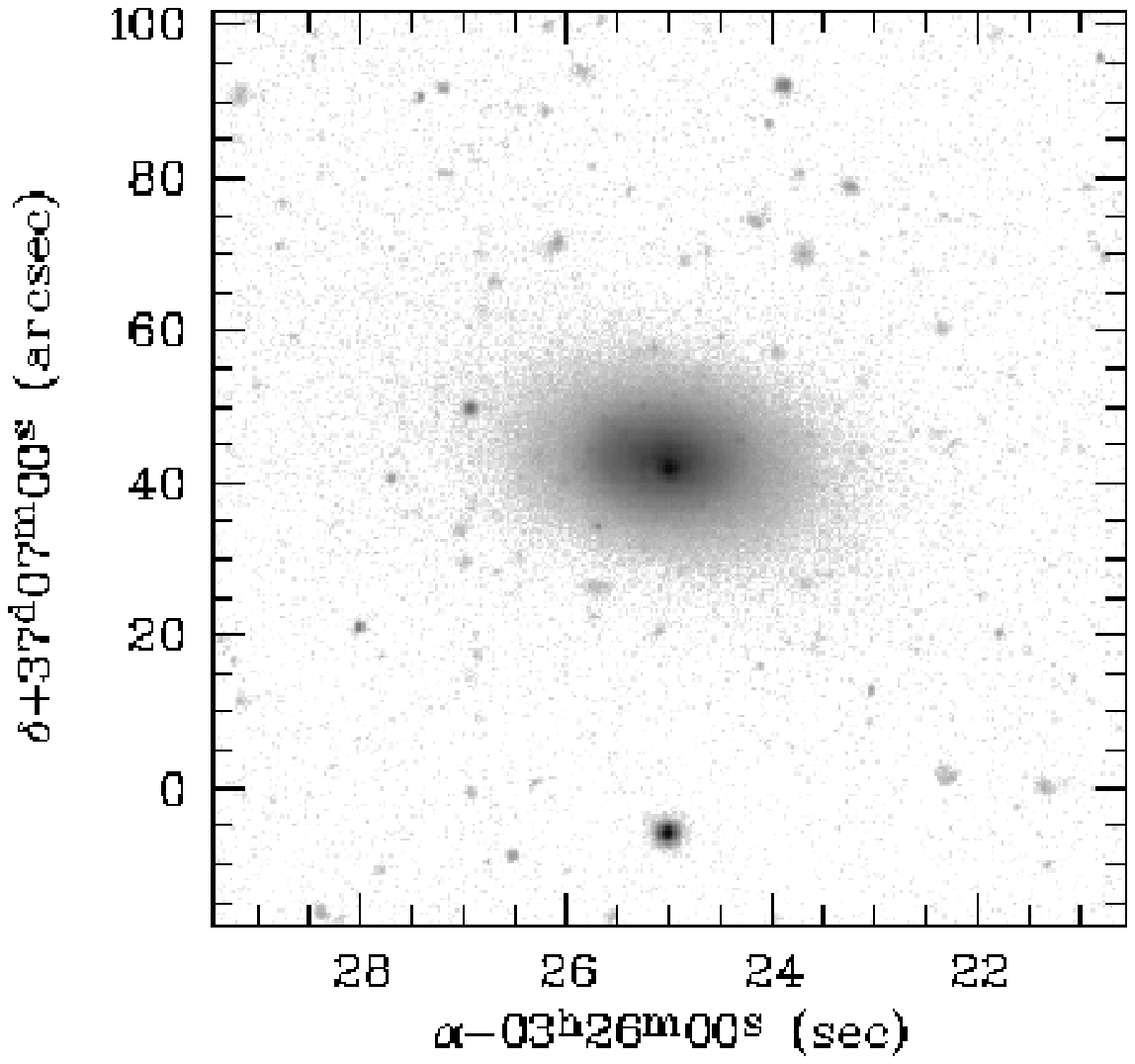"}
\caption{450~sec. B-image of the dE4 FCC046. The nucleus is offset by
1.1$''$ to the south-west of the center of the outer
isophotes. \label{FCC046_B}}
\vspace{8cm}
\special{hscale=70 vscale=70 hsize=700 vsize=230 
         hoffset=-71 voffset=-240 angle=0 psfile="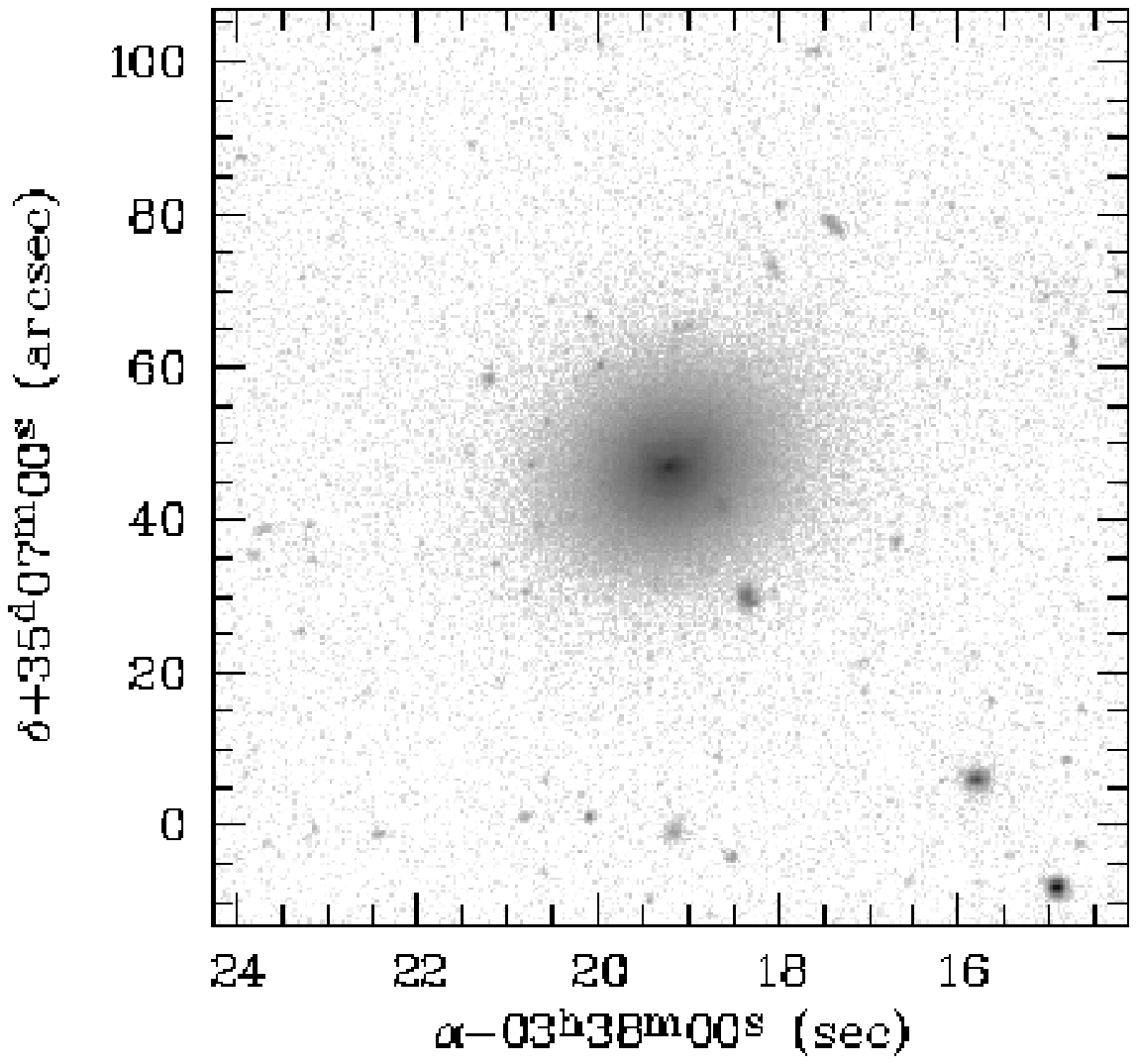"}
\caption{450~sec. B-image of the dE2 FCC207. \label{FCC207_B}}
\end{figure}

FCC046 (Fig. \ref{FCC046_B}) and FCC207 (Fig. \ref{FCC207_B})
(Ferguson \cite{fer}) were selected as targets for an ongoing ESO
Large Programme to study the structure and dynamics of dEs. These were
the only galaxies in our sample with published evidence of the
presence of an ISM and recent star-formation. Ionised hydrogen was
detected by Drinkwater {\em et al.} \cite{dri} in both galaxies. These
authors interpret this as photo-ionisation by young stars and use
Kennicutt's \cite{ken,ken2} calibration between the total SFR and the
H$\alpha+$[N{\sc ii}] equivalent width (EW)
\begin{equation}
{\rm SFR} \approx 2.7 \times 10^{-12}\, \frac{L_B}{L_{B,\odot}} {\rm EW(H}\alpha+{\rm [N{\sc
ii}]})\,M_\odot/{\rm yr},
\end{equation}
with $L_B$ and $L_{B,\odot}$ the B band luminosity of the galaxy and
the sun, respectively, to estimate the SFRs in these galaxies at $1-2
\times 10^{-3} M_\odot$/yr. Held \& Mould \cite{hel} present UBV
colors and metallicities of, amongst others, FCC207. They conclude
that FCC207 is too blue in U$-$B (U$-$B$=0.15$) and too metal-poor for
its B$-$V (B$-$V$=0.78$) and interpret this as a consequence of the
presence of a young stellar population. This motivated us to
investigate both objects more closely using BRI broad-band and
H$\alpha+$[N{\sc ii}] narrow-band imaging. In section \ref{odr}, we
discuss the details of the observations and data reduction. The B$-$R
color maps are presented in section \ref{bcm} and the results of the
H$\alpha+$[N{\sc ii}] narrow-band imaging are shown in section
\ref{hai}.

\section{Observations and data reduction} \label{odr}

The observations were carried out on 18 and 20 November 2001 with
Yepun ({\tt VLT}-{\tt UT4}) using {\tt FORS2}. We took 20 minute
exposures of FCC046 and FCC207 with the H\_Alpha/2500+60 filter
centered on 6604~{\AA} and with a FWHM$=64$~{\AA}. R band images
obtained during a previous run (1-8 November 2000) served as off-band
images. Two H$\alpha$-images of the spectrophotometric standard star
LTT9239 were taken for flux-calibration. During these observations,
the seeing (determined from the stars on the images) typically was
$0.7''-0.8''$~FWHM. The standard data reduction procedures (bias
subtraction, flatfielding, cosmic removal, interpolation over bad
pixels, sky subtraction) were performed with {\tt MIDAS}
\footnote{{\tt ESO-MIDAS} is developed and maintained by the European
Southern Observatory}. All science images were corrected for
atmospheric extinction (using the R band extinction coefficient~:~$k_e
= 0.13$) and interstellar extinction (we used the Galactic extinction
estimates from Schlegel {\em et al.} \cite{sch}~:~$A_R = 0.050$ for
FCC046 and $A_R = 0.039$ for FCC207). The images were finally
converted to units of electrons/second/pixel.

In order to find the correct scaling for the R-band images we adopted
the following strategy. The pure emission ``Em'' can be recovered from
a narrow-band image ``Nb'' and an R-band image ``Rb'' as
\begin{equation}
{\rm Em} = {\rm Nb} - ( c \times {\rm Rb} + \delta) \label{subtr}
\end{equation}
with $c$ the proper scaling constant and $\delta$ a correction for
possible faulty sky-subtraction. To find the best values for $c$ and
$\delta$, we first fitted the isophotes of the narrow-band and R-band
images in an annulus between $m_R=24.5$~mag/$\Box''$ and
$m_R=26.5$~mag/$\Box''$, which in retrospect did not contain any
emission (hence Em$=0$), using the standard {\tt MIDAS} {\tt FIT/ELL3}
command. Thus, a smooth version of this annulus could be constructed
for both images. The optimal $c$ and $\delta$ can be found by
minimising the expression $|{\rm Nb} - (c \times {\rm Rb} + \delta)|$
with Nb and Rb the smoothed versions of the annulus. With these values
in hand, the pure-emission image can be obtained using relation
(\ref{subtr}). $\delta$ was very small for both FCC046 and FCC207,
which makes us confident that the sky was properly subtracted in all
images. Since the H$\alpha$ and R-band overlap, subtracting an R-band
image in lieu of a continuum image entails a partial removal of some
H$\alpha+$[N{\sc ii}] light. The error thus introduced is of the order
of the ratio of the effective widths of the filters
(R-band~:~$W=165.0$~nm, H$\alpha$~:~$W=6.4$~nm), i.e. less than
4\%. Since this effect is negligible in comparison to the other
possible sources of error, we did not correct for it.

A pixel-value in the pure-emission image (corrected for both
atmospheric and interstellar extinction), denoted by $N_g$, expressed
in electrons/second, can be converted to flux units, $F'_g$, using the
formula
\begin{equation}
F'_g = N_g \times  \frac{\varphi_{\rm o}}{N_*}
  \int_0^\infty {\cal F}^*_\lambda(\lambda) \varphi_{\rm
  f}(\lambda)\,d \lambda\,\,\,\,\,{\rm W~m}^{-2}.
\end{equation}
Here, ${\cal F}^*_\lambda(\lambda)$ is the spectrum of a
flux-calibration standard star and $N_*$ is the measured flux of that
star, expressed in electrons/second. $\varphi_{\rm f}(\lambda)$ is the
transmission of the H$\alpha$ filter and $\varphi_{\rm o}$ the
transmission of the optics (which is basically constant for a
narrow-band filter). The prime on $F'_g$ indicates that this is the
flux incident on the CCD, after going through the telescope and
instrument optics and the narrow-band filter. This can also be written
as
\begin{eqnarray} 
F'_g &=& \varphi_{\rm o} \left[ F_{\rm{H}\alpha} \varphi_{\rm
f}(\lambda_{\rm{H}\alpha}) + F_{\rm{[N{\sc ii}]}_1} \varphi_{\rm
f}(\lambda_{\rm{[N{\sc ii}]}_1}) \right. \nonumber \\ && \left. \hspace*{4em} +
F_{\rm{[N{\sc ii}]}_2} \varphi_{\rm f}(\lambda_{\rm{[N{\sc ii}]}_2})
\right]
\end{eqnarray}
with $F_{\rm{H}\alpha}$, $F_{\rm{[N{\sc ii}]}_1}$ and $F_{\rm{[N{\sc
ii}]}_2}$ the incoming fluxes -- i.e. before going through the
telescope and instrument optics and the narrow-band filter -- of
respectively the H$\alpha$~6563\AA, the [N{\sc ii}]~6548\AA~and the
[N{\sc ii}]~6583\AA~emission line (approximated as
$\delta$-functions). This allows one to obtain the true incoming flux
of the H$\alpha$ emission line as
\begin{equation}
F_{\rm{H}\alpha} = \frac{ \frac{N_g}{N_*} \int_0^\infty {\cal
F}^*_\lambda(\lambda) \varphi_{\rm f}(\lambda)\,d \lambda} {
\varphi_{\rm f} ( \lambda_{\rm{H}\alpha} ) + \left[ \frac {F_{\rm{[N{\sc
ii}]}_1}} {F_{\rm{[N{\sc ii}]}_2}} \varphi_{\rm f} ( \lambda_{\rm{[N{\sc
ii}]_1}} )+\varphi_{\rm f}(\lambda_{\rm{[N{\sc ii}]_2}}) \right]  \frac {F_{\rm{[N{\sc
ii}]}_2}} {F_{\rm{H}\alpha}}}.
\end{equation}
The total incoming H$\alpha+$[N{\sc ii}] flux is simply
\begin{eqnarray}
F_{\rm em} &=& {F_{\rm{H}\alpha}} \left( 1 + \frac {F_{\rm{[N{\sc
ii}]}_1}}{F_{\rm{H}\alpha}} + \frac {F_{\rm{[N{\sc
ii}]}_2}}{F_{\rm{H}\alpha}} \right).
\end{eqnarray}
Since the H$\alpha$ filter is relatively flat-topped and the H$\alpha$
and [N{\sc ii}] lines are well inside the filter transmission curve,
the total flux is rather insensitive to the adopted relative
line-strengths. In the following, we will assume the mean value
$F_{\rm{[N{\sc ii}]}_2}/F_{\rm{[N{\sc ii}]}_1}=3$ for the ratio of the
line-strengths of the two Nitrogen lines (Macchetto {\em et al.}
\cite{mac}, Phillips {\em et al.} \cite{phi}). The ratio $F_{\rm{N{\sc
ii}}_2}/F_{\rm{H}\alpha}$ is not known and is treated as a free
parameter, varying between 0 and 2.
\begin{figure*}
\vspace{22cm}
\special{hscale=80 vscale=80 hsize=700 vsize=700 
         hoffset=0 voffset=-20 angle=0 psfile=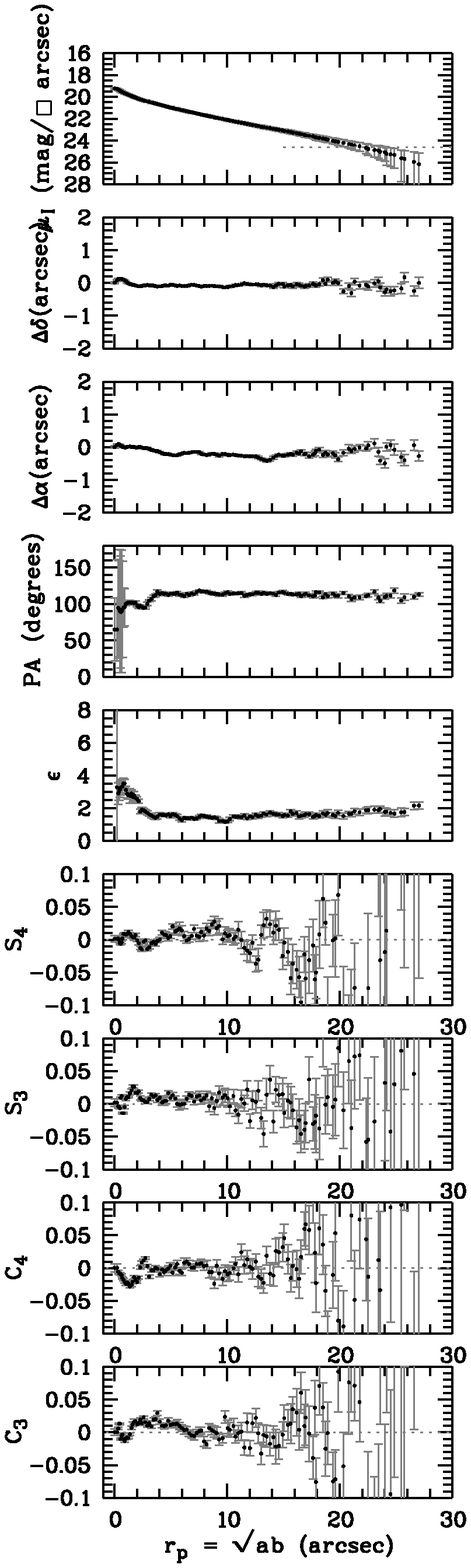}
\special{hscale=80 vscale=80 hsize=700 vsize=700 
         hoffset=200 voffset=-20 angle=0 psfile=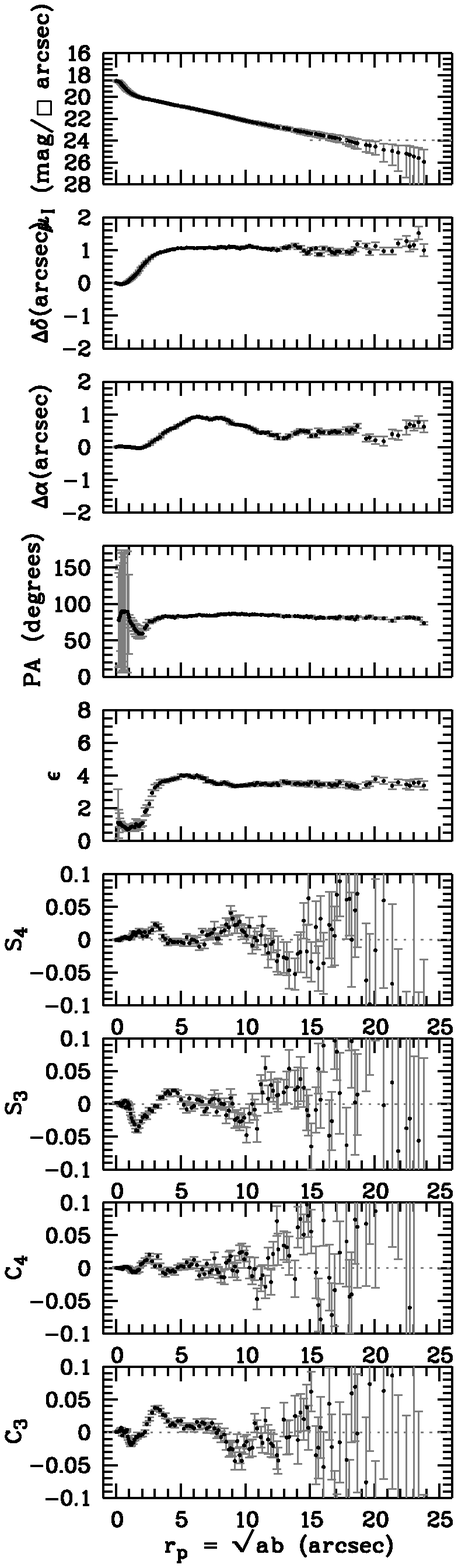}
\caption{Photometric properties of FCC207 (left) and FCC046 (right),
derived from the I band image, versus the geometric mean of the
semi-major and semi-minor distances $a$ and $b$. From top to
bottom~:~the I-band surface brightness $\mu_{\rm I}$ (the dotted line
corresponds to a surface brightness equal to 1\% of the sky level),
the deviation in declination $\Delta \delta$ and right ascension
$\Delta \alpha$ of the centers of the isophotes with respect to the
brightest point, the position angle PA, the ellipticity $\epsilon =
10(1-b/a)$, and the Fourier coefficients $S_4$, $S_3$, $C_4$ and $C_3$
that quantify the deviations of the isophotes from
ellipses. \label{surf_207}}
\end{figure*}
\begin{figure}
\vspace{8cm}
\special{hscale=70 vscale=70 hsize=700 vsize=230 
         hoffset=-71 voffset=-240 angle=0 psfile="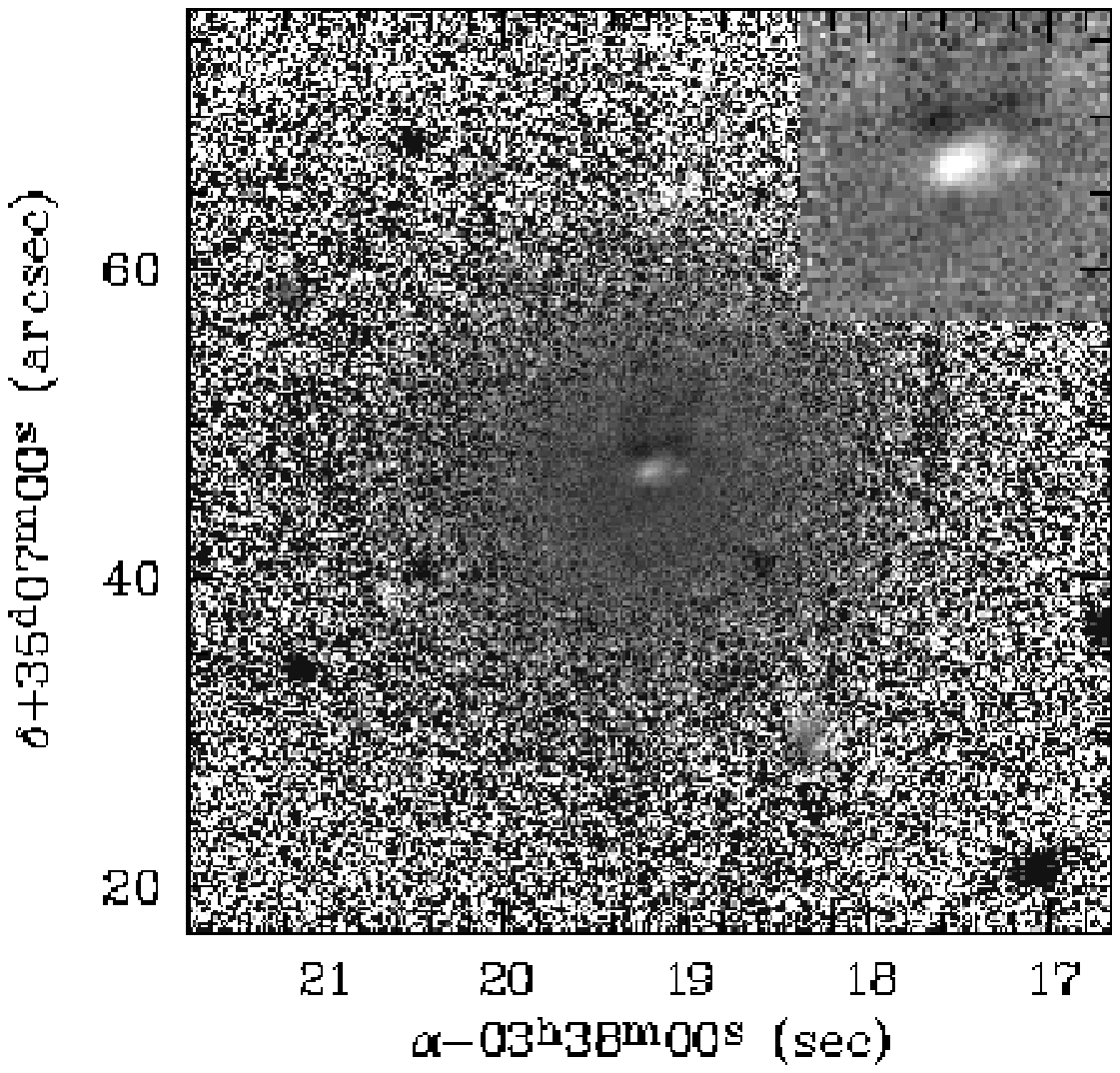"}
\caption{B$-$R color map of FCC207. The nucleus is rather blue
(B$-{\rm R}\approx 0.9$~mag) compared to the bulk of the galaxy
(B$-{\rm R}\approx 1.25$~mag). The inset shows the central $5'' \times
5''$ region with a different greyscale. To the north of the nucleus, a
signature of dust-absorption is visible ($\Delta({\rm B}-{\rm
R})=0.2$~mag). \label{FCC207_BR}}
\vspace{8cm}
\special{hscale=42.5 vscale=42.5 hsize=700 vsize=230 
         hoffset=0 voffset=240 angle=-90 psfile="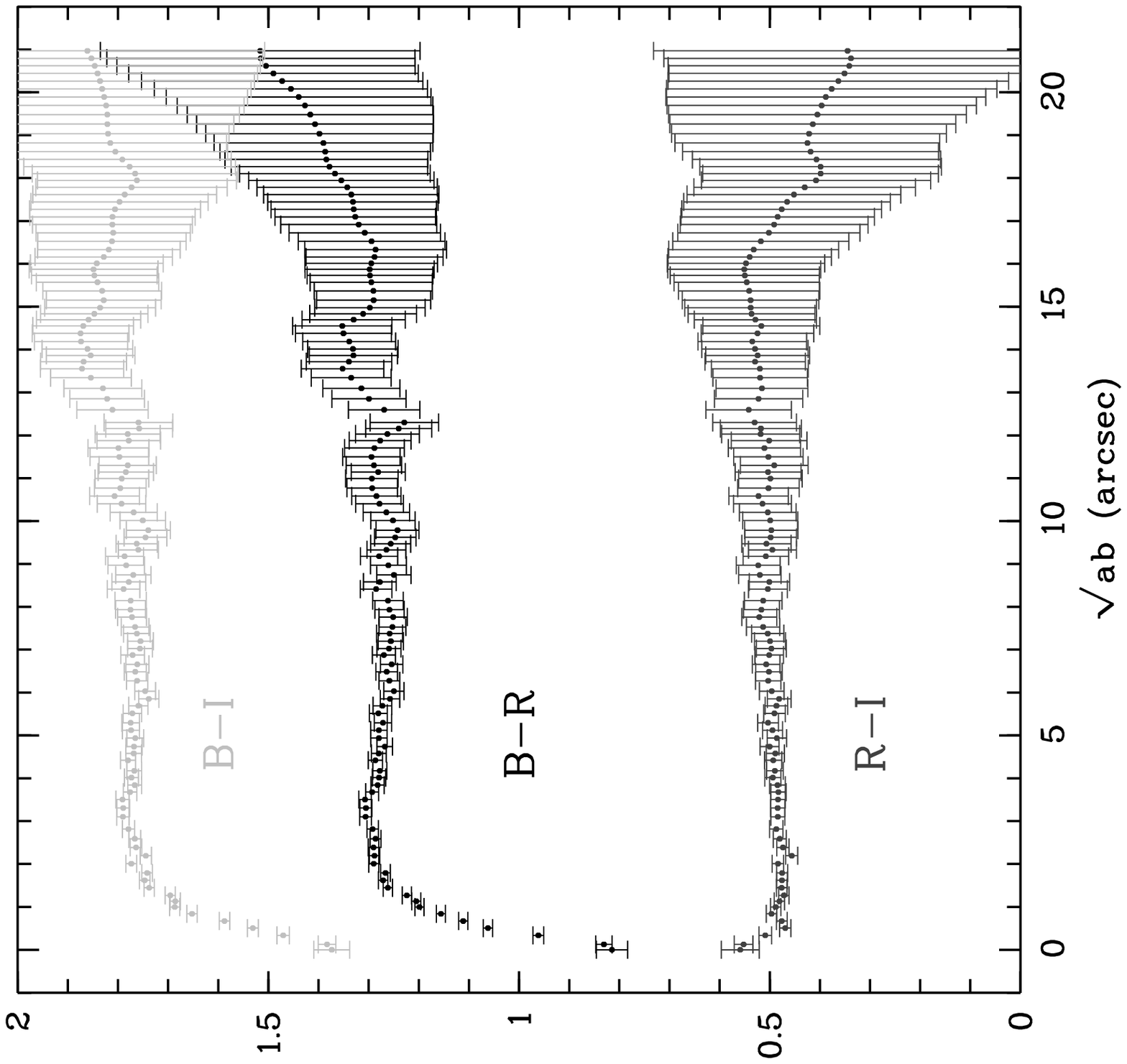"}
\caption{B$-$I, B$-$R and R$-$I profiles of FCC207 as a function of the
geometric mean of semi-major axis $a$ and semi-minor axis $b$
distance. Outside the nucleus, the colors are essentially 
constant. \label{FCC207_cm}}
\end{figure}
\begin{figure}
\vspace{8cm}
\special{hscale=70 vscale=70 hsize=700 vsize=230 
         hoffset=-71 voffset=-240 angle=0 psfile="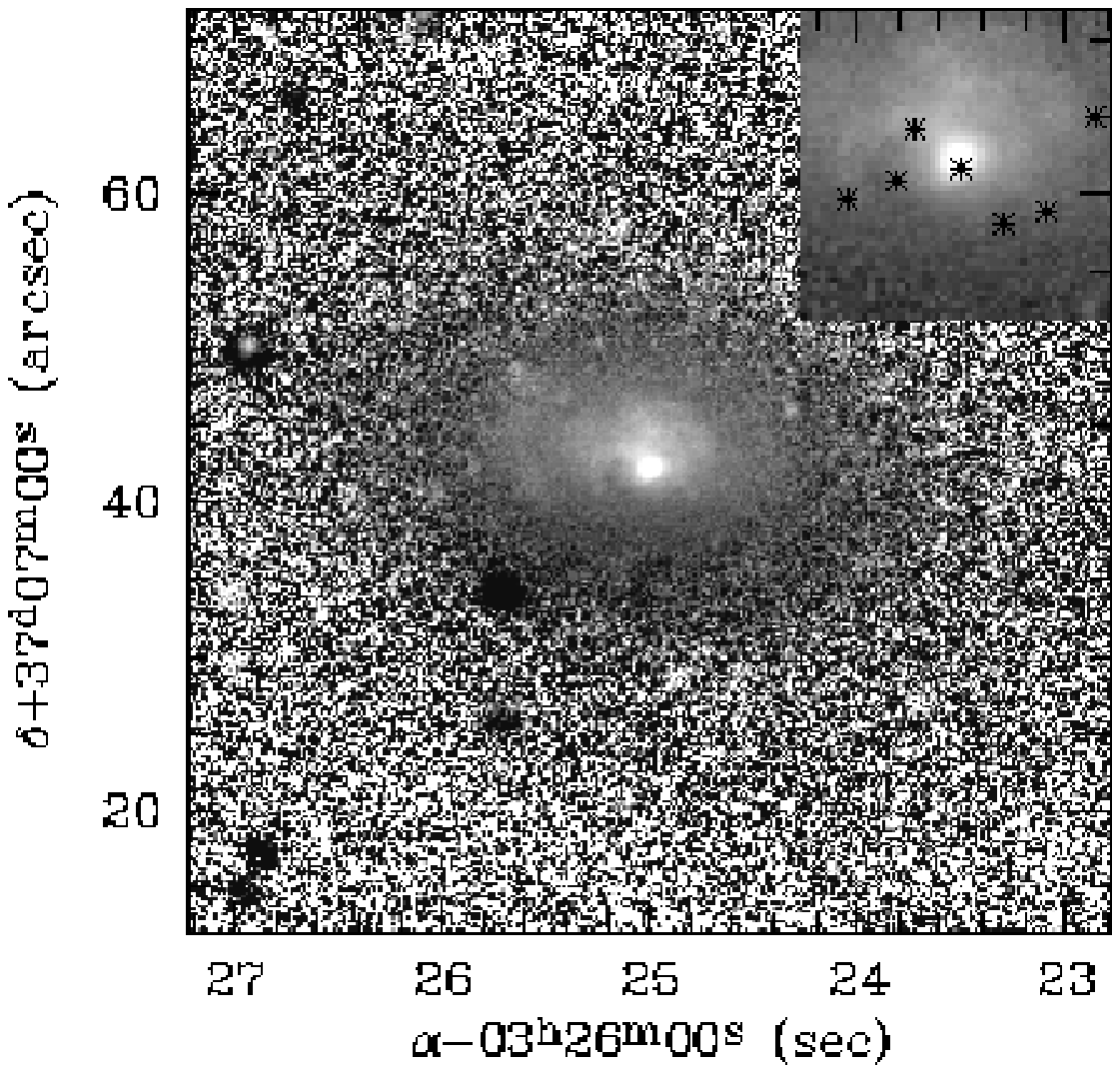"}
\caption{B$-$R color map of FCC046. White is blue, black is
red. Obviously, the off-center nucleus is very blue (B$-{\rm R}\approx
0.1$~mag). The inset shows the central $5'' \times 5''$ region with a
different greyscale. The asterisks mark the positions of the brightest
H$\alpha$ features (see section \ref{Hmass}).  \label{FCC046_BR}}
\vspace{8cm}
\special{hscale=42.5 vscale=42.5 hsize=700 vsize=210 
         hoffset=0 voffset=240 angle=-90 psfile="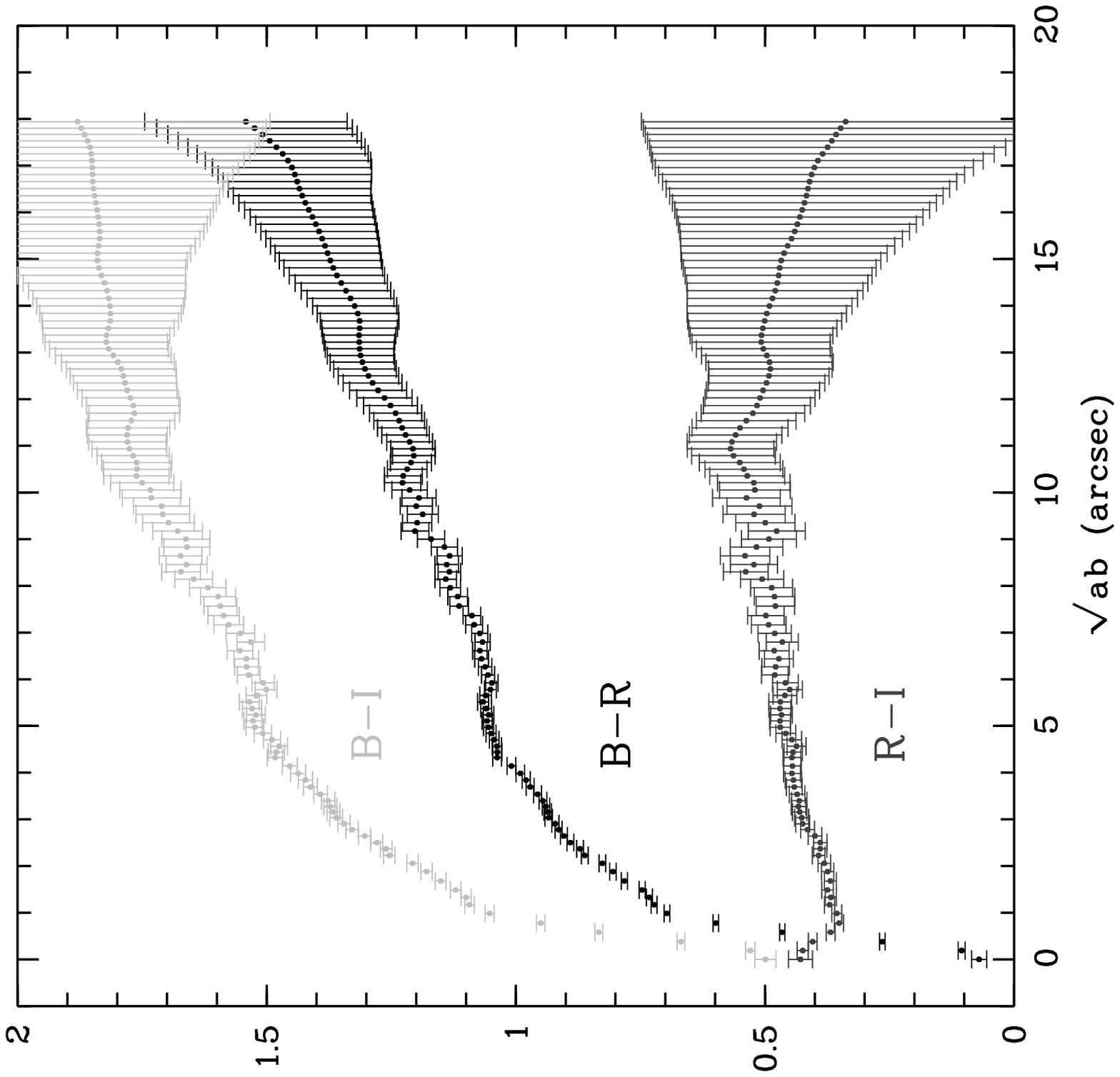"}
\caption{B$-$I, B$-$R and R$-$I profiles of FCC046 as a function of
the geometric mean of semi-major axis $a$ and semi-minor axis $b$
distance. The nucleus is clearly much bluer than the envelope. The
colors of the underlying stellar population become redder towards
larger radii. \label{FCC046_cm}}
\end{figure}

The rms scatter in the final pure-emission images is about
0.035~electrons/pixel/second (or $2.5 \times 10^{-20}\,{\rm W~m}^{-2}$
for $F_{\rm{[N{\sc ii}]}_2}/F_{\rm{H}\alpha}=1.38$, the average value
found by Phillips {\em et al.} \cite{phi} for a sample of normal
ellipticals).

\section{B$-$R color maps} \label{bcm}

The B, R and I images were used to extract surface brightness,
position angle and ellipticity profiles (see Figure
\ref{surf_207}). The deviations of the isophotes from a pure elliptic
shape were quantified by expanding the intensity variation along an
isophotal ellipse in a fourth order Fourier series with coefficients
$S_4$, $S_3$, $C_4$ and $C_3$~:
\begin{eqnarray}
I(\theta) &=& I_0 \left( 1 + C_3 \cos(3(\theta-{\rm PA}))+ C_4
\cos(4(\theta-{\rm PA})) + \right. \nonumber \\ && \left. S_3
\sin(3(\theta-{\rm PA}))+ S_4 \sin(4(\theta-{\rm PA})) \right)
\end{eqnarray}
with PA the position angle. All photometric parameters were fitted by
cubic splines as functions of semi-major axis distance. The galaxy
nucleus (i.e. the brightest pixel) was used as zeropoint for both $a$
and $b$, the semi-minor axis distance. This allowed to reconstruct the
surface brightness at a given point on the sky and to construct color
profiles (e.g. B$-$R as a function of radius).

\subsection{FCC207}

FCC207 has de-reddened magnitudes $m_I=14.39$~mag, $m_R = 14.86$~mag
and $m_B = 16.19$~mag (hence B$-{\rm R}=1.33$~mag, R$-{\rm
I}=0.47$~mag). Its nucleus has a distorted shape~: it is more
elongated than the bulk of the galaxy (E3 versus E2) and is somewhat
kidney-shaped. This is probably due to dust-absorption to the north of
the nucleus, noticeable in the B$-$R color map (Figure
\ref{FCC207_BR}) as a patch that is $\approx 0.2$~mag redder than its
surroundings. The nucleus (B$-{\rm R}=0.90$~mag) is significantly
bluer than the bulk of the galaxy (B$-{\rm R}=1.25$~mag). This
behavior is similar to what e.g. Bremnes {\em et al.} \cite{brem1}
find in dwarf galaxies in nearby groups. A small, slightly east-west
elongated blue object (B$-{\rm R}=1.10$~mag) can be seen to the west
of the nucleus. It is also visible in the H$\alpha$ image. Its
elongation rules out the possibility that it is a faint foreground
star. As can be seen in Figure \ref{FCC207_cm}, the B$-{\rm R}$,
B$-{\rm I}$ and R$-{\rm I}$ colors stay essentially constant outside
the nucleus. If a young stellar population is present outside the
nuclear region of FCC207 (the inner $2''$) then these stars are
apparently well mixed with the older population.

\subsection{FCC046}

\begin{figure*}
\vspace{13cm}
\special{hscale=70 vscale=70 hsize=700 vsize=430 
         hoffset=0 voffset=400 angle=-90 psfile="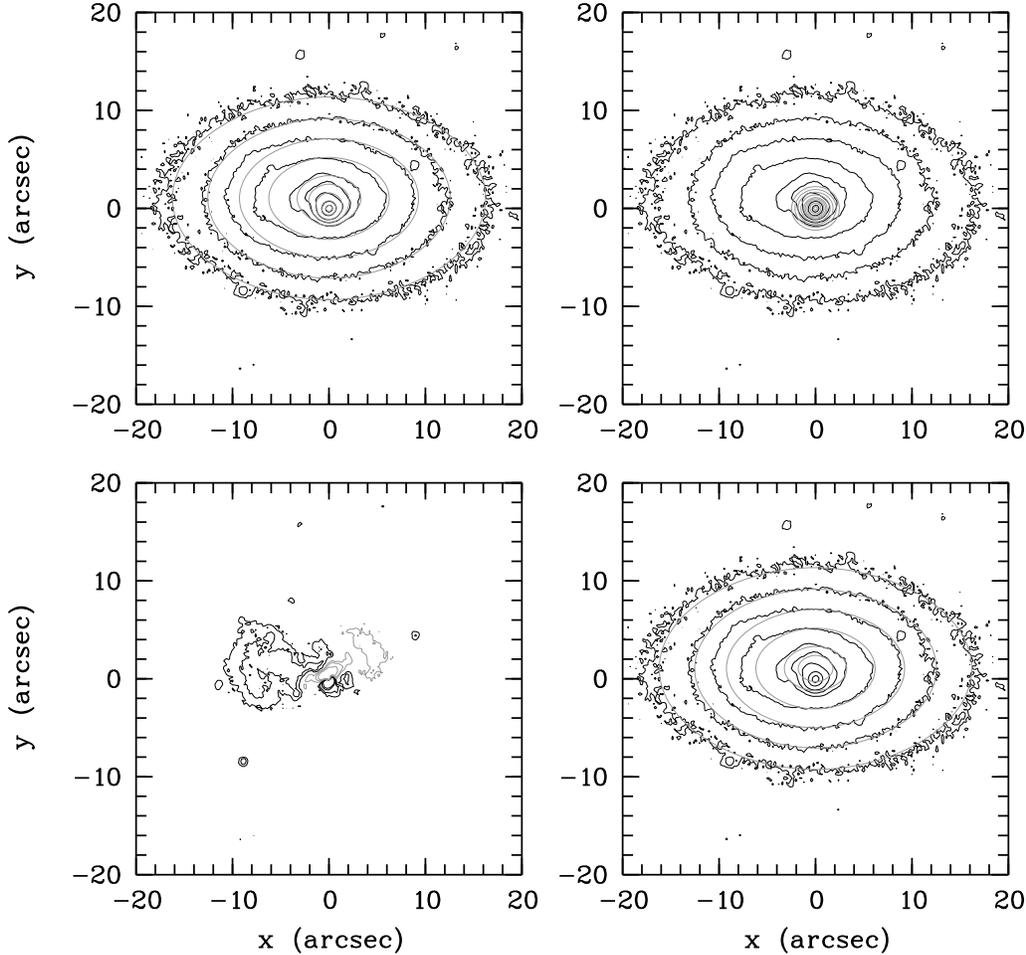"}
\caption{Two-component fit to the surface brightness of FCC046. The
galaxy is presented with the origin in the bright nucleus and the
$x$-axis along the major axis. The top left panel~:~isophotes of the
2-component model (grey) and the observed surface brightness (black);
top right panel~:~isophotes of the nucleus alone (grey); lower right
panel~:~isophotes of the axisymmetric envelope alone (grey). In all
these panels, the same isophotes are plotted. Lower left
panel~:~residue between the model and the data (black is positive,
grey negative). Apart from the off-center nucleus, FCC046 is obviously
also lopsided. \label{2comp}}
\end{figure*}

FCC046 is a rather blue object, with de-reddened magnitudes
$m_I=14.43$~mag, $m_R = 14.88$~mag and $m_B = 15.99$~mag (hence
B$-{\rm R}=1.11$~mag, R$-{\rm I}=0.45$~mag). The nucleus, a round (E0)
and blue (B$-{\rm R}=0.10$~mag) object (see Figure \ref{FCC046_BR}),
is offset by 1.1$''$ to the south-west of the center of the outer
isophotes (see Figure \ref{surf_207}). B$-{\rm R}$, B$-{\rm I}$ and
R$-{\rm I}$ color profiles are presented in Figure \ref{FCC046_cm} and
show a very different behavior than those of FCC207. The colors of the
stellar population become redder towards larger radii. The nucleus of
FCC046 is much bluer than those of nucleated dwarfs presented by
Bremnes {\em et al.} (e.g. \cite{brem1}). These authors typically find
$B-R \approx 0.5$ for the nucleus. The nucleus is resolved in the
B-band image. This implies that the nucleus is much larger than would
be expected for a typical dE. Even with the superior resolving power
of {\tt HST}, Lauer {\em et al.} \cite{lau} could not resolve the
nuclei of 5 nucleated Virgo dEs. The diameter (FWHM) of the nucleus
was estimated using the relation
\begin{equation}
{\rm FWHM}_{\rm true} = \sqrt{ {\rm FWHM}_{\rm obs}^2 - 
{\rm FWHM}_{\rm star}^2 } \label{fwhm}
\end{equation}
with ${\rm FWHM}_{\rm true}$ the true dimension, ${\rm FWHM}_{\rm
obs}$ its observed FWHM and ${\rm FWHM}_{\rm star}$ the average FWHM
of the stars in the image. The seeing, estimated from 10 stars in the
B-band image, was $0.82''\pm 0.04''$. The measured FWHM of the nucleus
is ${\rm FWHM}_{\rm obs}=1.1''$ or ${\rm FWHM}_{\rm true}\approx
65$~pc (for $H_0=75$~km/s/Mpc and a Fornax systemic velocity $v_{\rm
sys} = 1379$~km/s). We fitted a two-component model to the B-band
surface brightness of FCC046~:~an axisymmetric component centered on
the outer isophotes that represents the light of the underlying
stellar population and a round component centered on the position of
the nucleus. The results of this decomposition are presented in Figure
\ref{2comp}. The nucleus has a blue magnitude $m_B = 18.55$~mag
($M_B=-12.77$) and comprises about 10\% of the total B-band luminosity
of the galaxy. It should be noted that the nucleus of FCC046 was
apparently not visible on the photographic plates on which Ferguson's
catalog \cite{fer} was based, since it is classified as a dE4 (i.e. as
a non-nucleated dwarf). The underlying stellar envelope deviates from
an axisymmetric mass model and shows a pronounced lopsidedness,
visible in Figure \ref{surf_207} as the bump in $\Delta \alpha$ in the
region $\sqrt{ab} \approx 2'' - 12''$. This asymmetry may be due to an
asymmetric distribution of few but bright young stars. This appears to
be plausible since the dynamical time scale, estimated as
\begin{equation}
\Delta t = \frac{2\pi r}{v_{\rm circ}} = \sqrt{\frac{ 16 \pi^2 r^3 }{ G M(r)
}} \approx 20~{\rm Myr}
\end{equation}
for typical values $r \approx 0.5$~kpc and $M(r) \approx 10^9
M_\odot$, is of the order of the life-time of the youngest stars so
these would not have had time to disperse all over the face of the
galaxy. The cause of persistent $m=1$ perturbations, that involve a
sizable fraction of a galaxy's mass, is still poorly
understood. Interactions are often invoked, especially in bright
galaxies, but examples of isolated lopsided galaxies are know
(particularly in H{\sc i}, Baldwin {\em et al.}  \cite{bal}). Since
there is no galaxy detected within a $20'\times 20'$ square centered
on FCC046, it seems unlikely that an encounter with another galaxy has
caused the lopsidedness. Dynamical instabilities have also been
invoked (Merritt \cite{mer} and references therein) but it remains
unclear whether such a hypothesis may work for all galaxies.

\section{H$\alpha$ imaging} \label{hai}

\subsection{The H$\alpha$ equivalent width}

Drinkwater {\em et al.} \cite{dri} have measured H$\alpha$ EWs of 108 confirmed
Fornax cluster members, including FCC046 and FCC207 with the {\tt
FLAIR-II} spectrograph on the UK Schmidt Telescope. The effective
aperture diameter of this system is at least 6.7$''$ (the fibre
diameter) and could be as large as 15$''$ (because of image movements
due to tracking errors and differential atmospheric refraction). They
find~:
\begin{eqnarray}
{\rm EW(FCC046)} &=& 2.1~{\rm \AA}, \nonumber \\
{\rm EW(FCC207)} &=& 2.2~{\rm \AA}. \nonumber 
\end{eqnarray}
For comparison, we calculated the EW inside some aperture radius $r$
from our images as~:
\begin{equation}
{\rm EW} = \frac{F_{\rm em}(r)}{F_{\rm cont}(r)} \Delta \lambda
\end{equation}
with $\Delta \lambda=64$~\AA~the FWHM of the redshifted H$\alpha$
filter and $F_{\rm em}(r)$ and $F_{\rm cont}(r)$ the total number of
counts inside a circular aperture with radius $r$ of respectively the
H$\alpha+$[N{\sc ii}] and the continuum image. We find~:
\begin{eqnarray}
{\rm FCC046} && \hspace*{-2em} \left\{ \begin{array}{ll}
                        \mbox{$F_{\rm em}=54.1\,$e$^-$/sec} & \\
			\mbox{$F_{\rm cont}(8'')=3745\,$e$^-$/sec} & \hspace*{-1em} \rightarrow {\rm EW}(8'')=0.9 \,{\rm \AA} \\
			\mbox{$F_{\rm cont}(3.5'')=1462\,$e$^-$/sec} & \hspace*{-1em} \rightarrow {\rm EW}(3.5'')=2.4 \,{\rm \AA},
                       \end{array} \right. \nonumber \\
{\rm FCC207} && \hspace*{-2em} \left\{ \begin{array}{ll}
                        \mbox{$F_{\rm em}=78.0\,$e$^-$/sec} & \\
			\mbox{$F_{\rm cont}(8'')=3300\,$e$^-$/sec} &  \hspace*{-1em}\rightarrow {\rm EW}(8'')=1.4 \,{\rm \AA} \\
			\mbox{$F_{\rm cont}(3.5'')=1250\,$e$^-$/sec} & \hspace*{-1em} \rightarrow {\rm EW}(3.5'')=3.7 \,{\rm \AA}.
                       \end{array} \right. \nonumber
\end{eqnarray}
Given the possible sources of error (photon shot-noise, sky and
continuum subtraction) that can affect our measurements, we consider
these values in good agreement with the EWs measured by Drinkwater
{\em et al.}  \cite{dri}.

\subsection{The H$\alpha+$[N{\sc ii}] and H$\alpha$ luminosities}

\begin{figure}
\vspace{8cm} \special{hscale=70 vscale=70 hsize=700 vsize=230
hoffset=-71 voffset=-240 angle=0 psfile="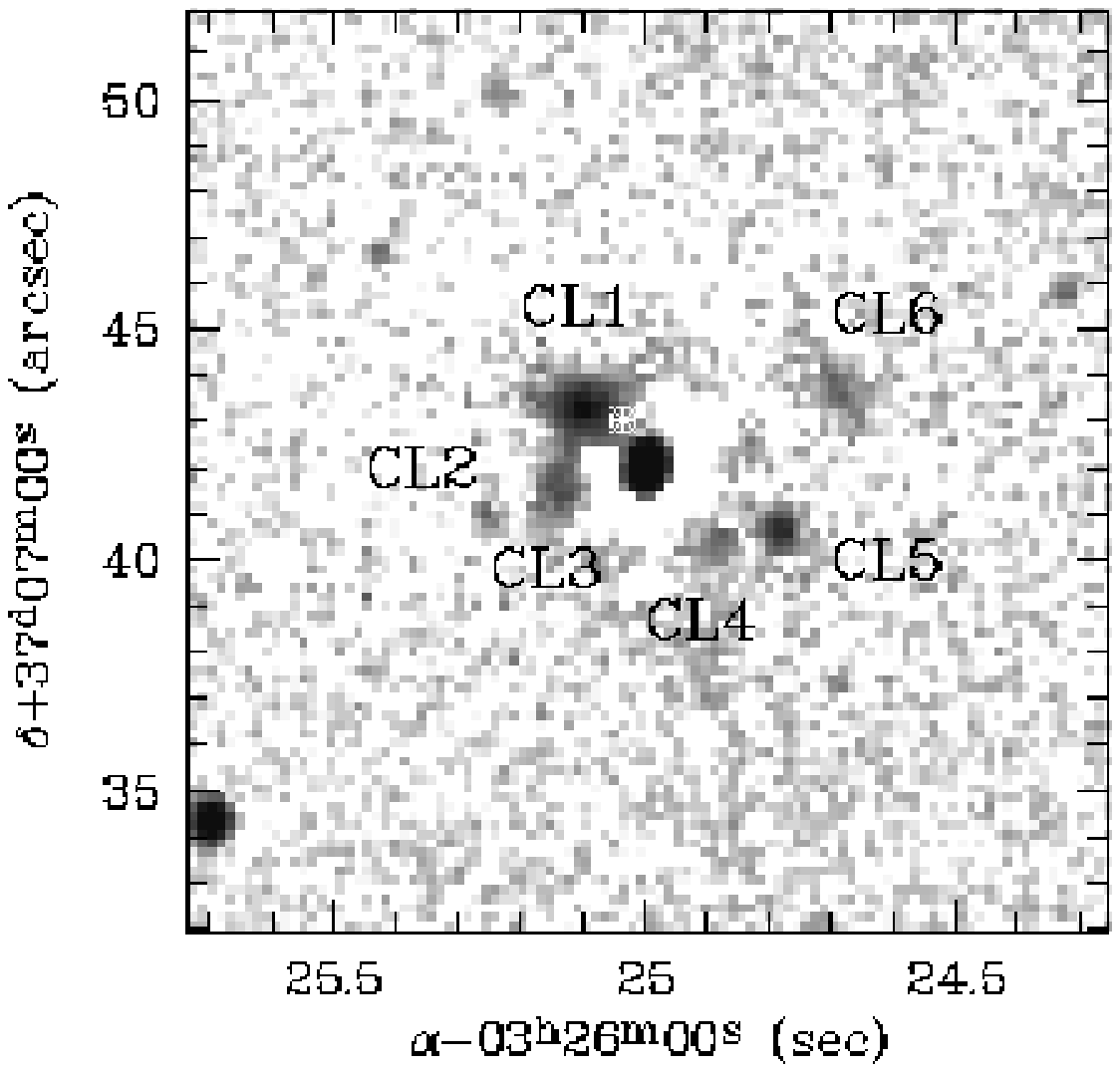"}
\caption{The pure emission image (H$\alpha$+[N{\sc ii}]) of FCC046. The
asterisk marks the center of the outer isophotes. The bright emission
feature in the center coincides with the off-center nucleus. The six
fainter emission ``clouds'' are labeled {\tt Cl1} up to {\tt
Cl6}. \label{FCC046_Ha}}
\vspace{8cm}
\special{hscale=70 vscale=70 hsize=700 vsize=230 
         hoffset=-71 voffset=-240 angle=0 psfile="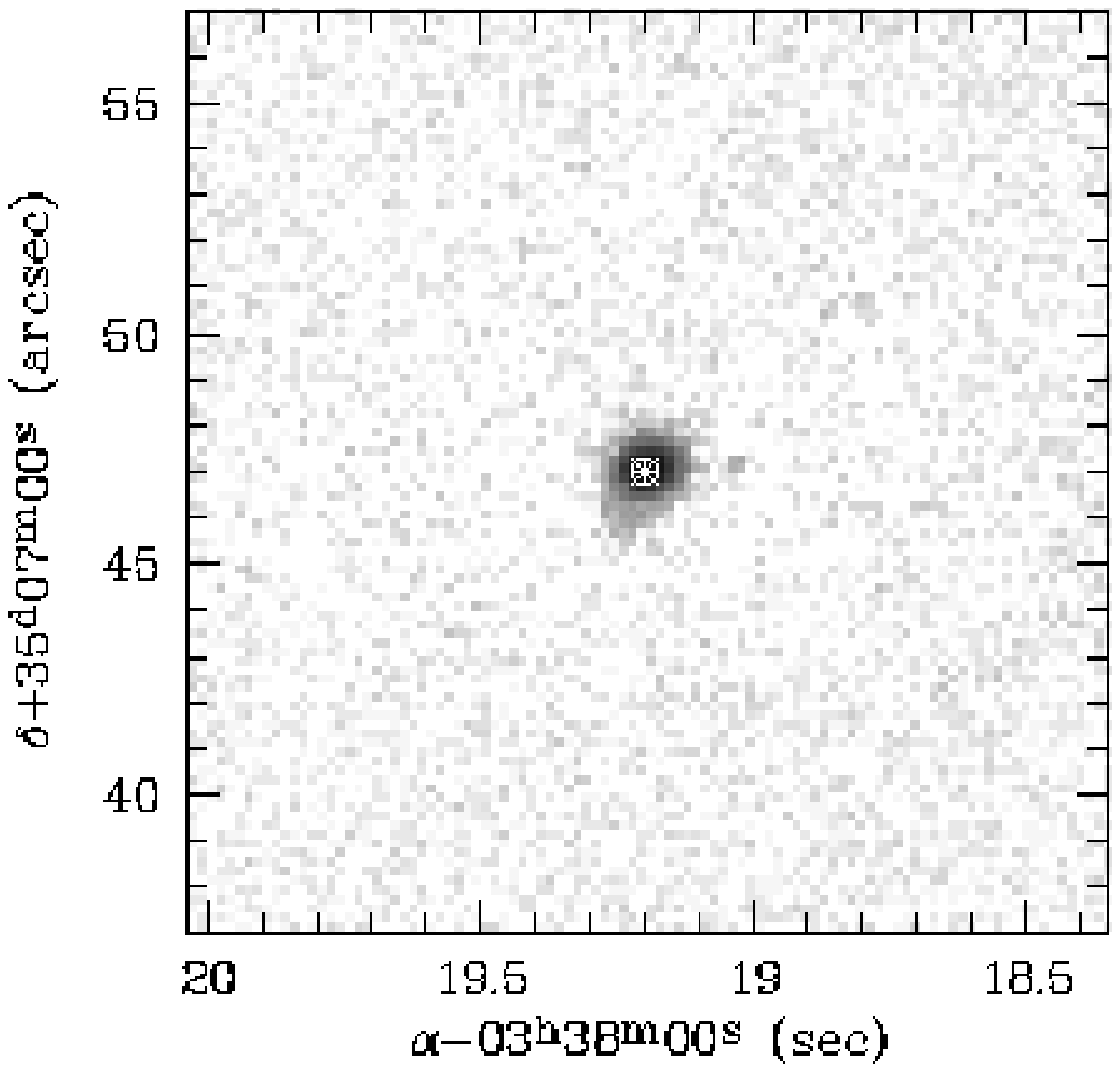"}
\caption{The pure emission image (H$\alpha$+[N{\sc ii}]) of FCC207.
The asterisk marks the center of the outer isophotes. A small emission
feature can be discerned 2$''$ to the west of the nucleus.
\label{FCC207_Ha}}
\end{figure}

\begin{figure}
\vspace{8cm}
\special{hscale=50 vscale=45 hsize=700 vsize=240 
         hoffset=-50 voffset=245 angle=-90 psfile="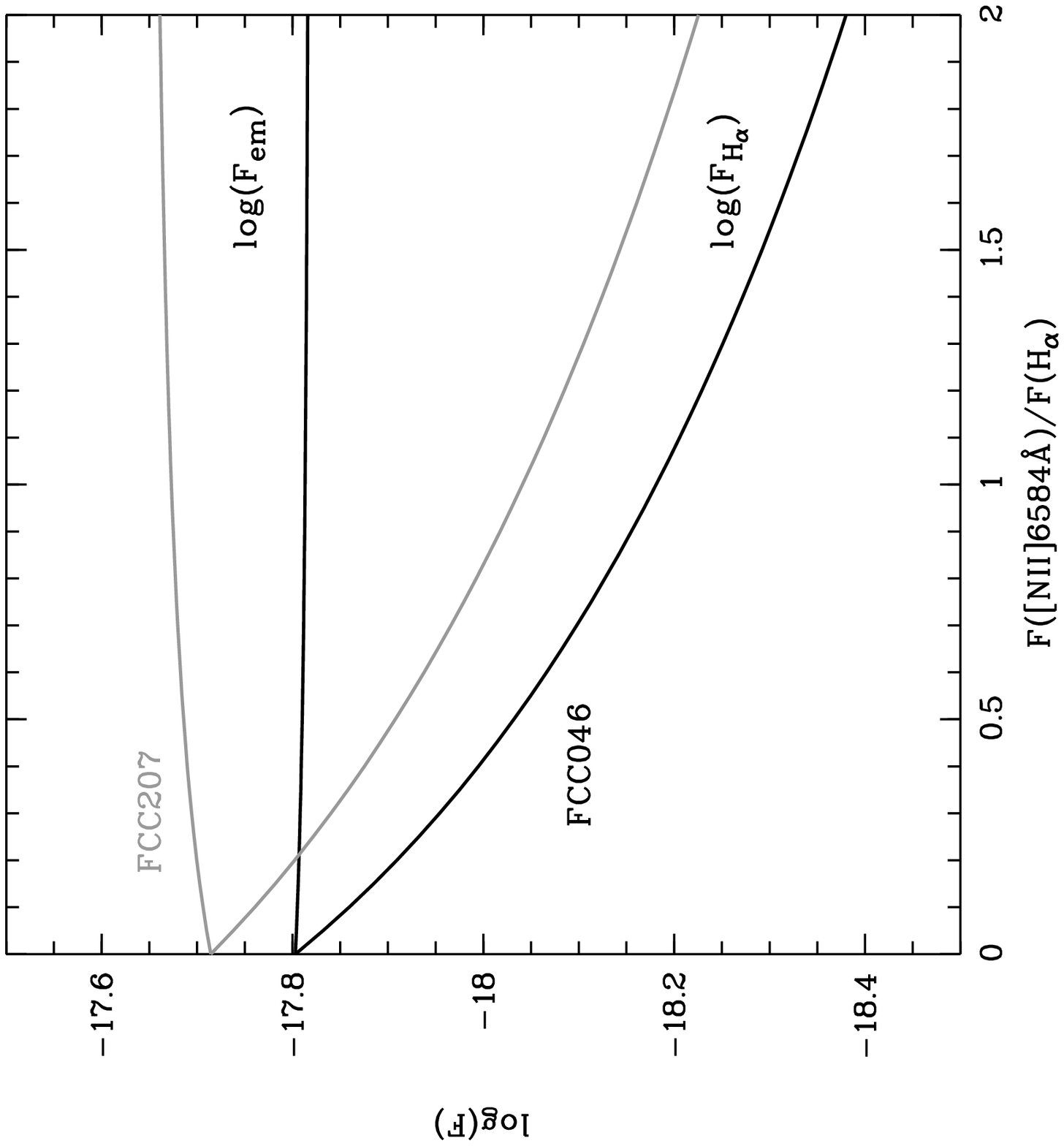"}
\caption{The logarithm of the total H$\alpha+$[N{\sc ii}] flux
($F_{\rm em}$) and the H$\alpha$ flux ($F_{\rm{H}\alpha}$) versus the
ratio of the strengths of the [N{\sc ii}]~8584\AA~and the
${\rm{H}\alpha}$ line. The total flux is virtually independent of this
line-ratio. \label{Fem2a}}
\end{figure}

\begin{figure}
\vspace{8.5cm}
\special{hscale=50 vscale=50 hsize=700 vsize=240 
         hoffset=-28 voffset=265 angle=-90 psfile="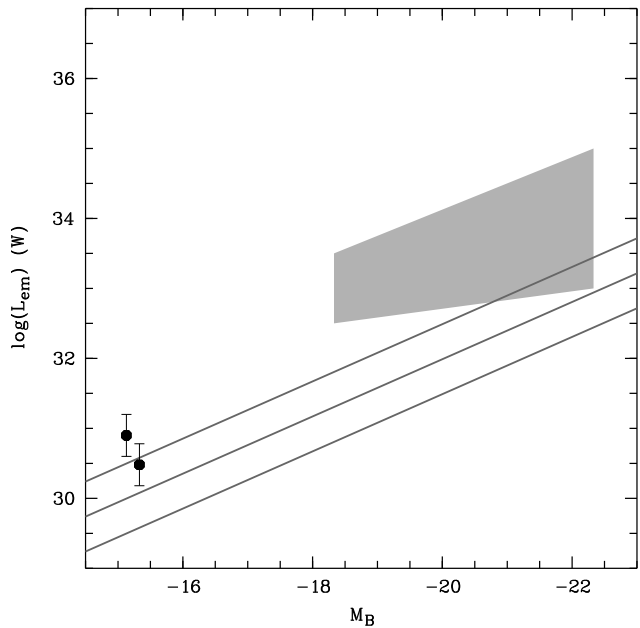"}
\caption{The total H$\alpha+$[N{\sc ii}] emission-line luminosity of
FCC046 and FCC207 versus absolute blue magnitude. The dark-grey lines
indicate the linear relation and its $1-\sigma$ deviation observed by
Phillips {\em et al.}  (1986). The Es and S0s observed by Buson {\em
et al.}  (1993) fill the light-grey area. All observations have been
converted to the distance scale adopted in this paper.
\label{buson}}
\end{figure}

Pure H$\alpha+$[N{\sc ii}] emission images of FCC046 and FCC207 are
presented in Figures \ref{FCC046_Ha} and \ref{FCC207_Ha}. For FCC046,
we find $F_{\rm em}({\rm FCC046}) = 1.53 - 1.57 \times 10^{-18}\,{\rm
W~m}^{-2}$, corresponding to a total luminosity $L_{\rm em}({\rm
FCC046}) = 6.21 - 6.37 \, h_{75}^{-2} \times 10^{30}\,{\rm W}$.  The
range of values is given for $F_{\rm{[N{\sc
ii}]}_2}/F_{\rm{H}\alpha}=0-2$ (see Figure \ref{Fem2a}). The central
emission peak comprises about half of the luminosity. It alone has a
luminosity of about $3 \times 10^{30}$~W. The total flux of FCC207 is
somewhat higher~:~$F_{\rm em}({\rm FCC207}) = 1.93 - 2.18 \times
10^{-18}\,{\rm W~m}^{-2}$, which yields a total luminosity $L_{\rm
em}({\rm FCC207}) = 7.83 - 8.84 \, h_{75}^{-2} \times 10^{30}\,{\rm
W}$. These numbers can be compared to those found by Buson {\em et
al.}  \cite{bus}, Kim \cite{kim}, Phillips {\em et al.}  \cite{phi}
and Shields \cite{shi} for normal elliptical and S0 galaxies. The
luminosities of the central emission peaks in FCC046 and FCC207 are
compared to those of ellipticals in Figure \ref{buson}. Typical
emission luminosities for these galaxies lie in the range $L_{\rm em}
= 10^{33} - 10^{35}$~W, i.e. more than a 1000 times brighter. The fact
that the luminosity of the nuclear emission in these dEs agrees fairly
well with the trend of normal Es -- extrapolated over more than 2
magnitudes -- suggests that the ionising mechanism, at least for the
central emission, is the same and therefore somehow related to the
stellar population.

The total H$\alpha$-flux of FCC046 is $F_{{\rm H}\alpha}=4.17 - 15.7
\times 10^{-19}\,{\rm W~m}^{-2}$, depending on the value of
$F_{\rm{[N{\sc ii}]}_2}/F_{\rm{H}\alpha}$. This translates into a
total H$\alpha$ luminosity $L_{{\rm H}\alpha}=1.69- 6.37 \,
h_{75}^{-2} \times 10^{30}\,{\rm W}$, about half of which is emitted
by the central peak corresponding to the galaxy's nucleus. The total
H$\alpha$-flux of FCC207 is somewhat higher~:~$F_{{\rm
H}\alpha}=5.95-19.3\times 10^{-19}\,{\rm W~m}^{-2}$, corresponding to
$L_{{\rm H}\alpha}=2.41 - 7.83 \, h_{75}^{-2} \times 10^{30}\,{\rm
W}$. Binette {\em et al.} \cite{bin} propose photo-ionisation by
post-AGB stars as a source for the central emission in elliptical
galaxies. Using their prescriptions, we derive central H$\alpha$
luminosities of the order of $2 \times 10^{30}$~W, i.e. comparable to
what is observed. Hence, blindly interpreting the central H$\alpha$
emission as evidence for star-formation may be somewhat audacious. We
can however check our results and use Kennicutt's \cite{ken}
calibration between the total SFR and the H$\alpha$ luminosity,
\begin{equation}
{\rm SFR} \approx 8.93 \times 10^{-35} L_{{\rm H}\alpha} E_{{\rm H}\alpha}
\end{equation}
with $E_{{\rm H}\alpha}=1$~mag the internal extinction factor. We
obtain
\begin{eqnarray}
{\rm SFR(FCC046)} &=& 0.4 - 1.4 \times 10^{-3} M_\odot/{\rm yr} \nonumber \\
{\rm SFR(FCC207)} &=& 0.5 - 1.8 \times 10^{-3} M_\odot/{\rm yr},
\end{eqnarray}
in good agreement with the estimates based on the EWs given by
Drinkwater {\em et al.} \cite{dri}.

\subsection{H{\sc ii} masses} \label{Hmass}

\begin{figure}
\vspace{8.25cm}
\special{hscale=48 vscale=48 hsize=700 vsize=260 
         hoffset=-20 voffset=260 angle=-90 psfile="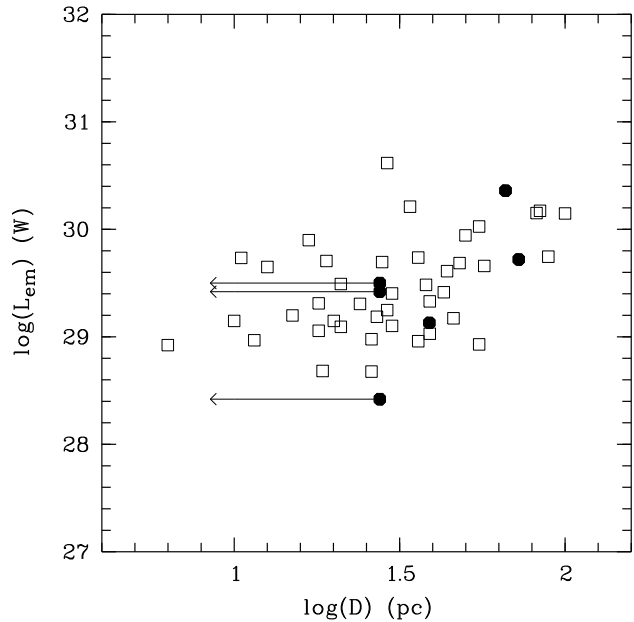"}
\caption{H$\alpha+$[N{\sc ii}] luminosity of supernova remnants in M33
(empty squares) versus their diameter. The six clouds identified in
FCC046 are presented as black dots. Three clouds have diameters that
are too small to be measured (i.e. smaller than $D \approx 30$~pc)
under the given seeing conditions. \label{SNR}}
\end{figure}

The total mass in ionised hydrogen can be written as
\begin{equation} 
M_{{\rm H}{\sc ii}} = \frac{L_{{\rm H}\alpha}}{4 \pi j_{{\rm
H}\alpha}} m_{\rm H} N_e
\label{mhy}
\end{equation}
with $L_{{\rm H}\alpha}$ the total H$\alpha$ luminosity, $m_{\rm H}$
the mass of the hydrogen atom and $N_e$ the electron density in the
gas. The hydrogen H$\alpha$ emissivity $j_{{\rm H}\alpha}$ is given by
\begin{equation} 
4 \pi j_{{\rm H}\alpha}=N_e^2 \alpha_{{\rm H}\alpha} h \nu_{{\rm
H}\alpha} = 3.544 \times 10^{-32} N_e^2 \, {\rm W~cm}^{-3} \label{e1}
\end{equation}
in ``case B'' recombination, i.e. complete re-absorption of all Lyman
photons in an optically thick nebula (Osterbrock \cite{ost}, Spitzer
\cite{spi}, Macchetto {\em et al} \cite{mac2}). Each Lyman photon
emitted from a level with $n \ge 3$ is later on converted to (a)
Balmer photon(s) plus one Lyman $\alpha$ photon, thus raising the flux
in the Balmer lines. The production coefficient $\alpha_{{\rm
H}\alpha}$ (calculated for $T=10^4$~K) is insensitive to the electron
density (it changes by only 4\% if $N_e$ is raised from 1~cm$^{-3}$ to
10$^6$~cm$^{-3}$) and varies as $T^{-0.8}$ as a function of
temperature. Using equations (\ref{mhy}) and (\ref{e1}), the ionised
hydrogen mass can be written concisely as~:
\begin{eqnarray}
  M_{{\rm H}{\sc ii}} &=& 23.72 \, \left( \frac{ 1000 \, {\rm
        cm}^{-3}}{N_e} \right) \left( \frac{L_{{\rm H}\alpha}}{10^{30}
        \rm W} \right) M_\odot \nonumber \\ &=& 2.85 \left(
        \frac{ 1000 \, {\rm cm}^{-3}}{N_e} \right) \left( \frac{
        F_{{\rm H}\alpha}}{10^{-23} {\rm W~cm}^{-2}} \right) \times
        \nonumber \\ && \hspace{5em}\left( \frac{r}{10~{\rm Mpc}}
        \right)^2 M_\odot, \label{mhy2}
\end{eqnarray}
cf. Kim \cite{kim}. In the following, we will assume the value $N_e =
1000$~cm$^{-3}$ for the electron density to be in accord with most
other authors and to be able to directly compare our ionised hydrogen
masses with the literature (however, Spitzer \cite{spi} advocates $N_e
= 100$~cm$^{-3}$ as a typical value for both Galactic H{\sc ii}
regions with diameters of the order of 100~pc and for supernova
remnants). Using equation (\ref{mhy2}), the mass of the ionised
hydrogen gas in FCC046 can be estimated at $M_{{\rm H}{\sc ii}}
\approx 40 - 150 \,M_\odot$ and at $M_{{\rm H}{\sc ii}} \approx 60 -
190 \, h_{75}^{-2} \,M_\odot$ in FCC207.

\subsection{FCC046~:~a starforming dE?}

The ${{\rm H}{\sc ii}}$ emission of FCC046 is distributed over a
bright central region and six fainter clouds, labeled {\tt Cl1} to
{\tt CL6} in Figure \ref{FCC046_Ha}. {\tt Cl1}, {\tt Cl2}, {\tt Cl5},
and {\tt Cl6} are identifiable in the B$-$R color map. {\tt Cl1} and
{\tt CL6} are part of the bluish nebulosity to the north of the
nucleus whereas {\tt Cl2} and {\tt Cl5} show up as individual blue
spots, about 0.1~mag bluer than their immediate surroundings. The
diameters (FWHM) of these clouds were estimated using equation
(\ref{fwhm}). We fitted gaussian profiles to 11 stars in the
pure-emission image of FCC046 and found ${\rm FWHM}_{\rm star} =
0.78'' \pm 0.06''$. Hence, clouds with an observed FWHM smaller than
0.84$''$ (or a diameter smaller than $\approx 30$~pc) cannot be
regarded as resolved. Clouds {\tt Cl1}, {\tt Cl3}, and {\tt Cl6} are
resolved under the given seeing conditions. In Table \ref{tab1}, the
diameters and luminosities of the clouds are listed. In the fourth
column, we give the emission rate of hydrogen ionising photons $Q_{\rm
max}$ needed to produce the luminosity $L_{\rm em}$ if the clouds
would be H{\sc ii} regions (i.e. we assume that all the light is in
the H$\alpha$ line to obtain an upper limit for $Q$)~:
\begin{equation}
Q_{\rm max} = L_{\rm em} \frac{\alpha_B}{h \nu_{{\rm H}_\alpha} \alpha_{{\rm H}_\alpha}}
\end{equation}
with $\alpha_B=2.59 \times 10^{-13}$~cm$^3$s$^{-1}$ the ``case B''
recombination coefficient for $T=10^4$~K (Osterbrock \cite{ost}). An
upper limit for the diameter of a H{\sc ii} region, $D_{\rm
max}$, is then given by
\begin{equation}
D_{\rm max} = \left( \frac{ 6 Q_{\rm max} } { \pi N_e^2 \alpha_B } \right)^{1/3}.
\end{equation}
The values in Table \ref{tab1} are calculated for $N_e =
100$~cm$^{-3}$.
\begin{table}
\begin{tabular}{|c|c|c|c|c|} \hline
name 		& $\log(D)$	& $\log(L_{\rm em})$	& 	$\log(Q_{\rm max})$	&	$\log(D_{\rm max})$	\\\cline{1-5}
{\tt Cl1}  	&	1.87		&	30.36 		&	49.62		&	1.01		\\
{\tt Cl2}  	&	$<1.49$		&	28.52 		&	47.78		&	0.39		\\
{\tt Cl3}  	&	1.91		&	29.72 		&	48.98		&	0.79		\\
{\tt Cl4}  	&	$<1.49$		&	29.50 		&	48.76		&	0.72		\\
{\tt Cl5}  	&	$<1.49$		&	29.42 		&	48.68		&	0.69		\\
{\tt Cl6}  	&	1.64		&	29.13 		&	48.39		&	0.60		\\ \cline{1-5}
\end{tabular}
\caption{Second and third column~:~the logarithm of the measured
diameters $D$ (pc) and H$\alpha+$[N{\sc ii}] luminosities $L_{\rm em}$
(W) of the six emission clouds in FCC046. Fourth and fifth
column~:~upper limits for the logarithm of the ionising photon
emission rates $Q_{\rm max}$ (s$^{-1}$) and diameters $D_{\rm max}$
(pc) if the clouds would be H{\sc ii} regions with
$N_e=100$~cm$^{-3}$. \label{tab1}}
\end{table}
In Figure \ref{SNR}, the diameters ($D = \sqrt{ab}$ with $a$ and $b$
respectively the long and short axes FWHM) and H$\alpha+$[N{\sc ii}]
luminosities of M33 SNR remnants and of the six clouds identified in
FCC046 are presented (Long {\em et al.} \cite{lon}). The properties of
the largest clouds are consistent with those of SNRs. The luminosities
of the clouds are also compatible with those of H{\sc ii} regions
ionised by the light of single 05-B0 stars but at least {\tt Cl1},
{\tt Cl3} and {\tt Cl6} seem too large for this interpretation. Though
suggestive, this of course does not mean that they necessarily are
SNRs. Nebulae around Wolf-Rayet stars could be a plausible alternative
and are found in many irregulars and have appropriate luminosities and
diameters (Hunter~\&~Gallagher \cite{hun}, Chu~\&~Lasker \cite{chu}).
It is striking that these emission clouds, whatever their
interpretation, are not found predominantly inside the bluish
nebulosity to the north of the nucleus, something that would be
expected if the blue light is coming from a young population of
stars. Their true nature can of course only be assessed by
spectroscopy.

\section{Conlusions}

The similarities of the broad-band colours of FCC046 to those of
star-forming or amorphous dwarfs, its relatively strong core and the
presence of emission clouds support the conclusion that FCC046 is
actively forming stars, albeit at a very leasurely pace when compared
to BCDs and amorphous dwarfs who are about a factor 1000 more luminous
in H$\alpha$. The nuclear emission of FCC046 and FCC207 can be
adequately accounted for by photo-ionisation by post-AGB stars
although a contribution of H$\alpha$ emission from star-formation
cannot be excluded. Only the emission from the six clouds observed in
FCC046 (supernova remants, Wolf-Rayet nebulae) can be interpreted as
unambiguous evidence for recent or ongoing star-formation. The
presence of physically different emission regions makes the
interpretation of this emission in terms of a star-formation rate
cumbersome. High-resolution spectroscopy of a broad wavelength region
is required to measure the strengths of H$\alpha$, H$\beta$ and of
telltale O, N and S emission lines in the visible part of the
spectrum. These can be used as diagnostics to probe the physical
nature of the different emission clouds.
\begin{figure}
\vspace{6.25cm}
\special{hscale=38 vscale=38 hsize=700 vsize=180 
         hoffset=-15 voffset=210 angle=-90 psfile="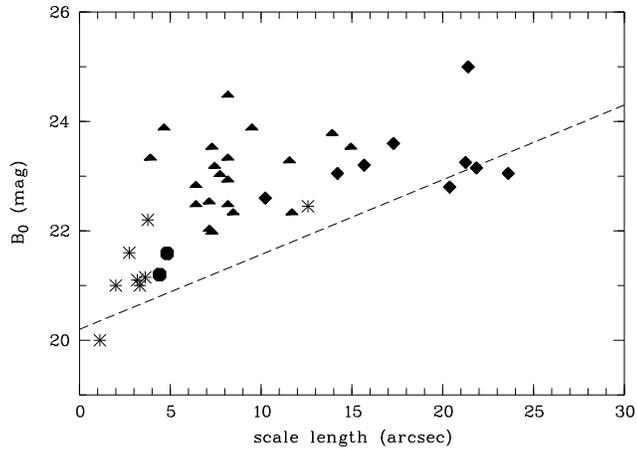"}
\caption{Extrapolated central B-band magnitude, $B_0$, versus
scale-length derived by fitting an exponential to the surface
brightness profile. Black dots~:~FCC046 and FCC207, asterisks~:~BCDs,
triangles~:~dEs, and diamonds~:~dIrrs.\label{drha}}
\end{figure}

Drinkwater {\em et al} \cite{dri} find no distinct class of
star-forming BCD galaxies but instead observe H$\alpha$ emission in
dwarf galaxies of all sizes and types. Among these, starforming dEs
like FCC046 may prove to be the descendents of more fiercely
star-forming dwarfs like BCDs which are not (or no longer) present in
Fornax. As a check, we fitted exponentials to the surface brightness
profiles of FCC046 and FCC207 and compared the extrapolated B-band
central surface brightnesses and the scale-lengths with those of the
Virgo dEs, BCDs, and dIrrs presented in Drinkwater~\&~Hardy
\cite{dha}. Both galaxies have scale-lengths in between those of BCDs
and dEs, and quite high B-band central surface brightnesses compared
to the dEs in the Drinkwater~\&~Hardy sample. These results support
our conjecture that dEs that contain ionised gas and possibly ongoing
low-powered star-formation can be considered as amissing link between
BCDs and traditional dEs.

\section*{Acknowledgments}

This research has made use of the NASA/IPAC Extragalactic Database
(NED) which is operated by the Jet Propulsion Laboratory, California
Institute of Technology, under contract with the National Aeronautics
and Space Administration. We thank the anonymous referee for helpful
comments. SDR wishes to thank Dr. Victor Debattista for useful
comments on the causes of lopsidedness. WWZ acknowledges the support
of the Austrian Science Fund (project P14783) and of the
Bundesministerium f\"ur Bildung, Wissenschaft und Kultur.

\end{document}